# Integrated membrane-free thermal flow sensor for silicon-on-glass microfluidics

*Vitaly V. Ryzhkov[1], Vladimir V. Echeistov[1,2], Aleksandr V. Zverev[1], Dmitry A. Baklykov[1,2], Tatyana Konstantinova[1], Evgeny S. Lotkov[1,2], Pavel G. Ryazantcev[1], Ruslan Sh. Alibekov[1], Aleksey K. Kuguk[1], Andrey R. Aleksandrov[2], Elisey S. Krasko[1], Anastasiya A. Barbasheva[1], Ilya A. Ryzhikov[1] & Ilya A. Rodionov[1,2]\**

[1] *FMN Laboratory, Bauman Moscow State Technical University, Moscow, Russia 105005.*

[2] *Dukhov Research Institute of Automatics, Moscow, Russia 127055.*

*email: irodionov@bmstu.ru; 105005, Moscow, 2nd Baumanskaya st., 5, building 1

**Abstract**

Lab-on-a-chip (LOC) forms the basis of the new-generation portable analytical systems. LOC allows the manipulation of ultralow flows of liquid reagents and multistep reactions on a microfluidic chip, which requires a robust and precise instrument to control the flow of liquids on a chip. However, commercially available flow meters appear to be a standalone option adding a significant dead volume of tubes for connection to the chip. Furthermore, most of them cannot be fabricated within the same technological cycle as microfluidic channels. Here, we report on a membrane-free microfluidic thermal flow sensor (MTFS) that can be integrated into a silicon-glass microfluidic chip with a microchannel topology. We propose a membrane-free design with thin-film thermo-resistive sensitive elements isolated from microfluidic channels and 4" wafers silicon-glass fabrication route. It ensures MTFS compatibility with corrosive liquids, which is critically important for biological applications. MTFS design rules for the best sensitivity and measurement range are proposed. A method for automated thermo-resistive sensitive elements calibration is described. The device parameters are experimentally tested for hundreds of hours with a reference Coriolis flow sensor demonstrating a relative flow error of less than 5% within the range of 2-30 µL/min along with a sub-second time response.

## 1. Introduction

The capabilities of modern analytical instruments are becoming insufficient given the ageing world population and numerous infectious disease outbreaks. Each of them, constitutes a potential risk of a new pandemic and requires on the spot rapid pathogen identification [1]. This problem can potentially be solved using microfluidic point-of-care devices (POC) capable of performing complex high-accuracy analyses [2-4]. Microfluidics allows handling nanoliters of liquid samples to measure ultra-low concentrations of analytes and to manipulate individual cells [5-9]. In healthcare, microfluidic devices can be useful for high-throughput screening due to process parallelization and automation. Combined with state-of-the-art plasmonic biosensors based on appropriate material science [10-12] providing up to single-molecule sensitivity, it opens up numerous opportunities for personalized medicine. Moreover, manipulating low volume samples and expensive chemicals on microfluidic devices requires high-precision control of system variables (volumetric flow rate, sampling volume *etc.*).

The most commonly used flow control devices for microfluidic systems are peristaltic pumps, syringe pumps and pressure control systems (PCS) with standalone flow sensors. Syringe pumps are widely used in microfluidics due to their ease of operation and ability to directly control volumetric flow rate [13]. However, such systems suffer from high response time and flow fluctuations [14]. Peristaltic systems are also convenient to use but low accuracy and high levels of flow pulsations significantly limit their field of application. Pressure control systems with standalone flow sensors provide high performance and smooth flow [15, 16]. However, connecting external flow sensors to a LOC requires supply tubes adding dead volume tens of times higher than the LOC's inner volume [16].

At the same time, the use of integrated flow sensors can help overcome these problems. A flow sensor is a compact device comprising a set of thin-film thermo-resistive elements, which are sensitive to the measured value associated with the flow, and an electric circuit, which recalculates the MTFS signal into the volumetric flow units [17]. Despite the variety of flow meter types (flowmeters based on pressure drop effects [18], on Coriolis effect [19], volumetric flowmeters [20], magnetic flowmeters [21], turbine flowmeters [22], ultrasonic flowmeters [23], etc.), the only thermal flow sensors (TFSs) are suitable for seamless integration in LOCs [24]. They exhibit a wide measurement range (from microliters to milliliters per minute), fast response (down to 100 ms), perfect measurement accuracy and are suitable for mass production using mature CMOS/MEMS technologies [17].

There is a number of TFS implementations have been introduced nowadays. In [25], the sensing elements of a device are in direct contact with the fluid making it impossible to work with corrosive biological fluids and reagents. Using a silicon wafer significantly reduces sensor's thermal inertia [26-27], which is necessary to perform accurate measurements of rapidly changing flows (greater than 20 μL/min/s). In addition, the complex fabrication process of the TFS`s membrane limits its integration into LOCs.

Moreover, the use of polymers (PDMS, PET film and other) [26,29] makes the device expendable and renders the technology unscalable. Therefore, there is a clear need for a microfluidic thermal flow sensor adapted for integration within microfluidic devices for aggressive biological fluids manipulation.

In this paper, we present a membrane-free microfluidic thermal flow sensor which can be integrated into a biocompatible silicon and glass microfluidic chip. The developed MTFS operates in calorimetric mode with a constant heater temperature. MTFS consists of a glass wafer with thin-film thermoresistive elements on the one side, and silicon wafer with channel and vias bonded to the other side of the glass. The sensing elements are not wetted with the liquid inside the channel, which allows using the sensor with corrosive biological fluids and under high pressures in LOCs. The channel cross-section is small as 400x50 μm and the MTFS experimental chip footprint presented here is 13x4 mm, but may be less than 3x4 mm. The technology is fully compatible with the mass production of such devices. We aim the MTFS for a wide range of cutting-edge applications, such as high-throughput automated microfluidic sample preparation for genomics, chromatographic chips, advanced LOCs, microfluidic devices for petrochemistry, microfluidic nanoemulsion generators.

## 2. Materials and methods

### 2.1. Simulations of heat and mass transfer in MTFS

The simulations were performed using Heat Transfer and Laminar Flow Modules in COMSOL Multiphysics software. The model was simulated in 3D and consisted of three parts: a planar glass wafer, a silicon wafer with a rectangular channel and water inside the channel. All physical properties were taken from the COMSOL materials library. To find an MTFS design with the best measuring performance, we varied the channel cross-section from 50x50 μm to 2000x200 μm and the glass wafer thickness. To define the best MTFS geometry, we tested each design under 0-80 μL/min flow rates and processed the raw data using MATLAB software.

*2.2. MTFS fabrication*

Elastomeric prototypes were fabricated using polydimethylsiloxane (PDMS) Sylgard 184 (Sigma-Aldrich, USA) and 500 μm thick double-side polished 4" UV-grade wafer (Siegert wafer GmbH, Germany). The master mold was fabricated from 525 μm thick <100> double-side polished 4" silicon wafer (Siegert wafer, Germany) and SU-8 permanent negative epoxy photoresist (Kayaku Advanced Materials, Inc., USA).

Silicon-on-glass devices were fabricated using a 525 μm thick <100> double-side polished 4" silicon wafer covered with 4 μm of wet $SiO_2$ on both sides (Siegert wafer, Germany) and 150 μm thick double-side polished 4" Borofloat 33 glass wafer (Schott AG, Germany). A Piranha solution ($H_2SO_4:H_2O_2$, 4:1) was used for removing organic contaminants (Technic France, France) at 120°C. Hexamethyldisilazane (HMDS) and buffered oxide etch (BOE) solution were used for adhesion promotion and silicon dioxide removal, respectively (Technic Inc., USA). For patterning functional layers, several photoresists and solvents were used: SPR220 and SPR955 photoresists and MF-CD-26 developer (DOW, Rohm&Haas Electronic Materials, USA); N-Methyl-2-pyrrolidone (NMP) ≥99,8%, VLSI Grade (Carl Roth GmbH + Co. KG, Germany); LOR 5A (MicroChem Corp., USA). Nickel pellets 99,995% and silicon dioxide pieces 99.99% were used for electron-beam evaporation of the sensing elements and the passivation layer (Kurt J. Lesker Company GmbH, Germany). Ethyl lactate solution (DOW, Rohm&Haas

Electronic Materials, USA), acetone (Technic Inc., USA) and isopropyl alcohol (Technic Inc., USA) were used for the diced MTFS chips cleaning before wedge bonding with Al-1% Si wire (Tanaka Precious Metals, Japan). Standard black 70 Shore EPDM seal rings 1.15x1 mm were used for leak-proof fluidic connection to the MTFS chip. The manifold was 3D printed with Grayscale Black Resin (Formlabs, USA).

*2.3. MTFS testing*

The experimental setup consisted of an experimental MTFS, a reference thermal flow sensor (MFS-3, Elveflow, France), a Coriolis flow sensor (mini Cori-flow M120, Bronkhorst, Netherlands), a multichannel pneumatic system (OB-1, Elveflow, France), a water feed reservoir, a flow resistor (F100, Dolomite Microfluidics, UK), a data acquisition system (DAQ6510, Keithley Instruments, USA), a PC and a drain tank.

A homemade LabVIEW program was used for real-time data visualization and processing following a preset algorithm.

## 3. Results and discussion

*3.1. MTFS design and operation principle*

There are 3 types of thermal flow sensors: thermoanemometric, calorimetric and time-of-flight. All three rely on the measurement of the local temperature field from heater(s) which is conjugated with moving media in the channel. The operating principle of thermoanemometric TFSs is based on the variation of the Nusselt number $Nu=q_c/q_d$ (where $q_c$ – convective heat flux, $q_d$ – conductive heat flux) of a heater working typically in a constant power or constant temperature mode. The working principle of calorimetric TFSs is based on the fact that a moving media causes asymmetry in the temperature field along the channel near the heater. The temperature field asymmetry is detected with the upstream and downstream thermal sensors equidistant from the heater. If thermophysical properties of the fluid are known, flow rate can be calculated from the TFS signal using a calibration curve. Time-of-flight TFSs measure the time interval it takes a heated flow section to reach the downstream thermal sensor located at a specified distance from the heater operating in an impulse mode [30,31].

The developed microfluidic sensor operates in the calorimetric mode with a constant heater temperature and the compensation of the ambient temperature fluctuations. The MTFS includes a microfluidic chip made from bonded silicon and glass wafers with a straight microfluidic channel between them (Fig. 1a). The channel has a rectangular cross-section and is formed in silicon. The thin-film thermoresistive heater $R_H$, the calorimetric sensors $R_{U1}$, $R_{U2}$, $R_{D1}$, $R_{D2}$ and the reference temperature sensors $R_{REF1}$, and $R_{REF2}$ (i.e. sensing elements) are located on the glass wafer outside of the channel. A huge advantage of our approach comes from the feasibility of multiple MTFS fabrication on a single microfluidic chip with a custom channel geometry. An example of the MTFS integration in a multichannel silicon-on-glass microfluidic device is shown in Fig. 1b.

The proposed calorimetric mode is implemented using three control circuits working simultaneously (Fig. 1c): the constant temperature difference (CTD) circuit, the voltage divider (VD) circuit and the measuring bridge (MB). The CTD circuit maintains a constant temperature difference between the heater and the reference temperature sensor, thereby compensating ambient temperature fluctuations and eliminating the risk of overheating on low flow rates.

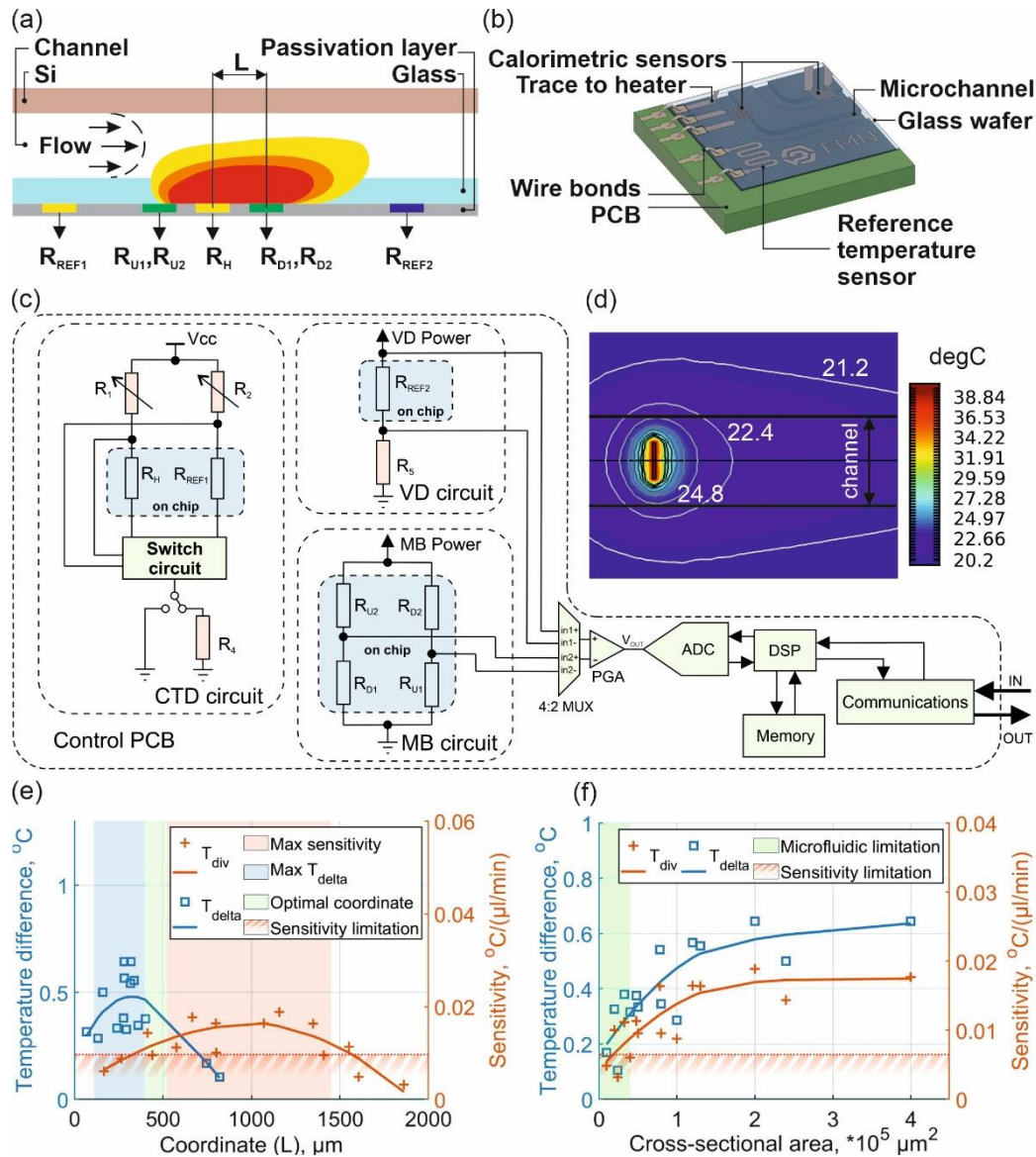

Figure 1 – (a) the MTFS operating principle scheme; (b) sectional view of a silicon-on-glass microfluidic device with integrated MTFSs; (c) the MTFS control PCB scheme; (d) temperature field distribution on the outer glass surface near the heater obtained from the simulation; (e) simulation results for MTFS with 150 μm thick glass wafer: graphs of the temperature difference between the upstream and downstream calorimetric sensors and the sensitivity of the MTFS, determining the coordinate **L** of the calorimetric sensors; (f) determining the channel dimensions to meet both «microfluidic» and «sensitivity» limitations.

*December 21, 2022*

The VD circuit includes an independent reference temperature sensor for on-chip temperature measurements, which ensures stable and reliable flow measurements when the temperature in the room vary significantly (19-30 ºC). The MB implements a Wheatstone bridge measurement method using calorimetric sensors.

The thermal output signals in the CTD circuit are defined by $V_H$ and $V_{REF1}$ voltages across the on-chip resistors $R_H$ and $R_{REF1}$, respectively. Depending on the voltage difference between $V_H$ and $V_{REF1}$, the output of the operational amplifier (OP-AMP) is either "high" (equal to VCC) or "low" (equal to zero). These "logic levels" open or close the MOSFET ("high" = open, "low" = close). Compared to conventional circuits [1], where a "source mode" switch with a BJT was implemented, the CTD circuit used here implements a "sink mode" switch with a MOSFET. When the MOSFET is closed, the current through the $R_H$ thermistor is small due to the large resistance of the thermistor $R_4$. This drastically reduces the heater power to near zero value (< 0,1 uW). When the MOSFET is open, the current through the $R_H$ resistor increases, thus maintaining the desired temperature difference between the liquid flow and the chip.

The output signal of the on-chip MB circuit is a differential voltage on inputs *in2+* and *in3-* of the multiplexer (4:2 MUX), which is then amplified using a programmable-gain amplifier (PGA) and converted to the $V_{OUT}$ value by the analog-to-digital converter (ADC). The $V_{OUT}$ directly depends on the temperatures of the calorimetric sensors which are associated with the flow rate in the channel. Assuming equality of the calorimetric sensors resistance values at the same temperature $R_{D1} = R_{U1} = R_{D2} = R_{U2}$ as well as their Temperature Coefficients of Resistance (TCR), the MB output signal can be calculated using the following equation:

$$V_{OUT} = \frac{GR_0 I_2 \alpha (T_D - T_U)}{2},$$

where G is the gain of signal amplification; $\alpha$ is the TCR of on-chip calorimetric sensors; $T_D$ is the temperature of $R_{D1}$ and $R_{D2}$, $T_U$ is the temperature of $R_{U1}$ and $R_{U2}$; $I_2$ is the amperage of the current source on the ADC IC; $R_0$ is the resistance of each calorimetric sensor at a temperature of $T_0$.

To get an on-chip local ambient temperature value, the VD circuit was implemented. The VD circuit output signal is the voltage on the $R_{REF2}$ sensor (inputs *in0+* and *in1-* of the ADC IC) measured in a constant current mode ($I_1$ source). The VD signal is then amplified by PGA and digitalized in ADC. The resulting MTFS signal from the MB is recalculated to get the flow value by the digital signal processing (DSP) unit taking into account the acquired VD circuit signal.

*3.2. Simulation results*

To get the best MTFS accuracy and sensitivity, a multivariate sensor design optimization was carried out. In the simulation, heat and mass transfer effects in the MTFS under low-Re conditions were studied. We varied the width and depth of a channel, glass wafer thickness and fluid flow velocity (an exhaustive parameter list is shown in Supplementary, Table S1). The developed multivariate optimization model was intended to define three main MTFS design parameters: width (W) and depth (D) of a channel, as well as the distance between the heater and the calorimetric sensors (L). Assuming that we can measure the temperature only on the outer glass surface (Fig.1d) and the biggest temperature field asymmetry can be found across a line on this surface along the channel, this line was set as the region of interest for further data processing.

Temperature difference ($T_{delta}$) in the region of interest between upstream ($R_U$) and downstream ($R_D$) calorimetric sensors depends on the flow rate and is directly associated with the MTFS signal magnitude. In turn, the change in the $T_{delta}$ value caused by a given flow alteration ($T_{div}$) is indicative of sensor sensitivity and can be defined as the partial derivative of $T_{delta}$ concerning the flow rate variable. These two parameters sufficiently represent the MTFS performance for the simulation, so we used them as criteria of optimization, the larger the better. The minimum sensor sensitivity threshold was set to 0.095 °C/(µL/min) ("sensitivity limitation"), and the upper limits of microfluidic channel width and depth were set to 500 µm and 80 µm, respectively ("microfluidic limitation").

We formulated the MTFS design rules for maximum sensitivity and accuracy based on hundreds computational experiments, which were performed and analyzed. As shown in Fig. 1e, to achieve the highest sensitivity, the distance L must be in the range of 250-500 µm, where the trade-off between the maximum $T_{delta}$ and $T_{div}$ is achieved. Based on the simulation results and the aforementioned limitations (Fig. 1f), a design with a 400x50 µm channel was selected for fabrication. However, it is worth noting that the best performance for the considered non-microfluidic designs could be achieved with the channel cross-sections of 1000x200 µm or 650x120 µm. Glass thickness decrease from 150 µm to 100 µm increases sensitivity on an average of 70% within the selected flow range (shown in the Supplementary, Fig. S1).

### 3.3. MTFS fabrication route

We started with standard PDMS technology to prove the MTFS concept. PDMS soft lithography is a common fabrication method that has been well-described elsewhere [32]. Using PDMS technology we fabricated and tested the general technical solutions, several versions of the MTFS, as well as various microfluidic chips, including a chip for accurate reagents dilution with two independent MTFSs (Fig. S2a). PDMS is a brilliant prototyping material for fast and cheap microchannel fabrication with leak-proof $O_2$ plasma bonding. However, it is impossible to perform long-term flow measurements using jelly-like material. The main shortcomings of PDMS hindering precise and reliable flow measurements are its softness and susceptibility to temperature changes, causing the inconsistent cross-sectional area of the channel, low chemical resistance, gas permeability and low durability.

Here, we focused on a silicon-on-glass technology. One of the key features of the proposed thermal flow sensor is the possibility to fabricate it together with advanced silicon-on-glass LOCs. The technology allows mass production of sensors, obtaining hundreds of items from a single wafer, which significantly reduces the prime cost. The MTFS consists of two parts: a microfluidic part containing channels, and a sensory part, which involves a chip with sensing elements. The main fabrication steps are shown in Fig. 2a.

The microfluidic part was fabricated using DRIE through a two-step hard mask. The mask was patterned using two rounds of photolithography (Fig. 2a, steps 1-5). Anodic bonding was used to close the channel. The thermoresistive elements for the sensory part were patterned on glass using lift-off photolithography. The same process was used for thermoresistive elements passivation (Fig. 2a, steps 6-10). Electric and fluidic interfaces were implemented using wire bonding and assembly with a manifold (Fig. 2a, steps 11-12). An explicit explanation of the fabrication technology is provided in the Supplementary materials.

Along with the LOC-integrated MTFS (Fig. S2b), we fabricated a standalone MTFS (Fig. S2c) with a serial interface and a display to indicate the flow rate value.

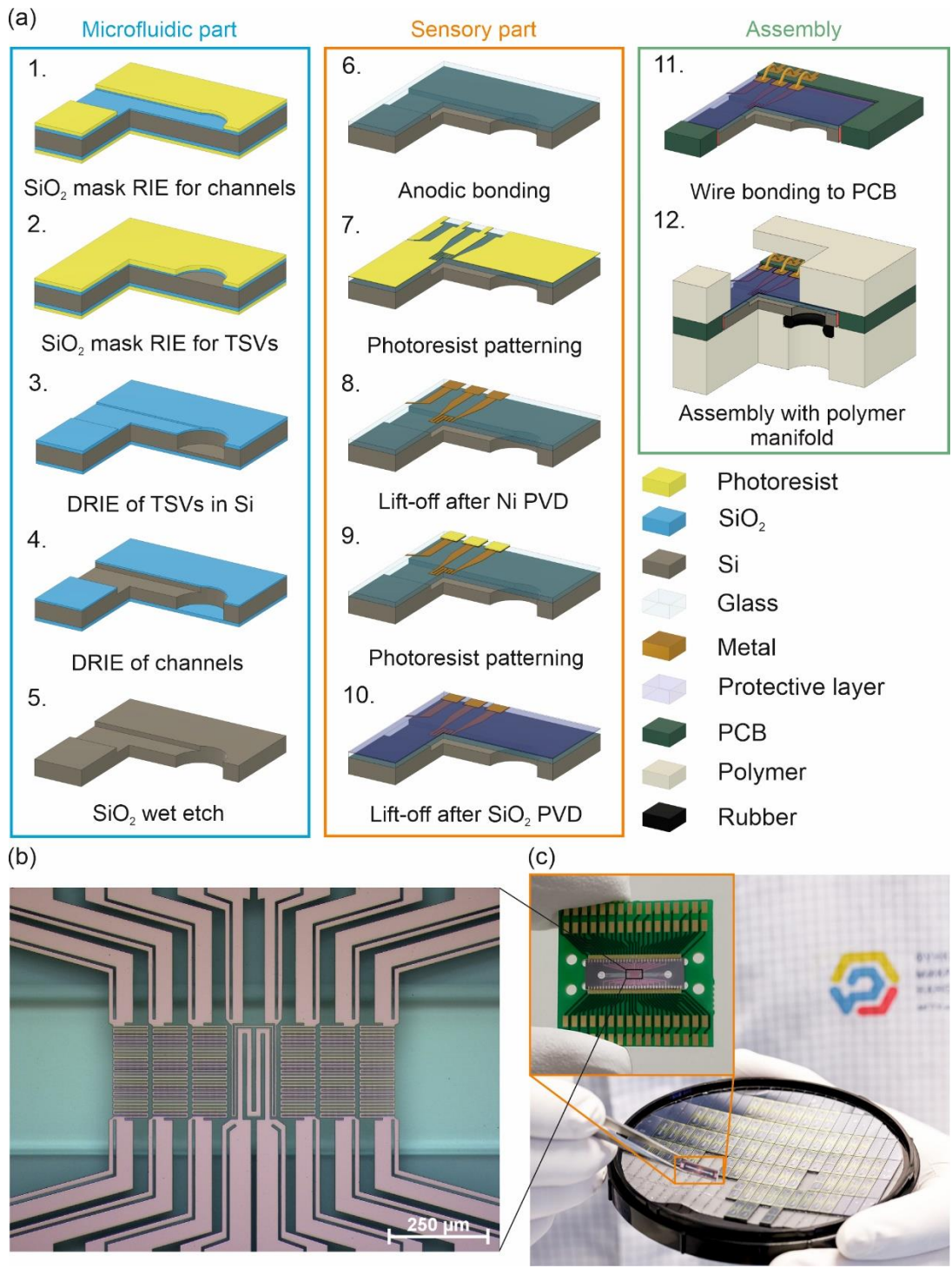

Figure 2 – (a) the main steps of the MTFS fabrication technology, including fabrication of a microfluidic part in silicon, of a sensory part on a glass wafer and device assembly; (b) micrograph of one of the tested MTFS; (c) a diced wafer with MTFSs and a magnified view of the MTFS wired on a PCB.

*December 21, 2022*

*3.4. MTFS characterization*

For the MTFS experimental testing we developed an automated calibration method aimed to define its calibration curve, absolute and relative accuracy, working range and response time. The equipment used in the experimental setup is described in Section 2.3. The pressure controller connected through a flow resistor was used to control the flow rate through three subsequently connected flow sensors (Fig. 3a). The two reference flow sensors (thermal and Coriolis) measured the actual flow rate in the common channel. The Coriolis flow sensor was used as the main reference sensor because of its well-known highest accuracy. A similar experimental scheme has been widely used for flow meter characterization [29]. A flow resistor was installed to accurately set low flow rates and suppress oscillations. Raw data from all the three flow sensors were synchronized in the LabVIEW program for real-time visualization, analysis, elimination of irregular measurements obtained under transient flow conditions, saving the filtered data and polynomial interpolation of the calibration curve.

To experimentally characterize the MTFS, we programmed the pressure controller to change the flow stepwise with each period of constant flow lasting 20 sec. The flow rate and MTFS signal were captured using the data acquisition system (analog, voltage) and PC (digital, serial port) during the stationary flow periods (Fig. 3b). A single calibration procedure included 3-20 full-range stairstep-like passages and took 12-40 minutes. It is worth saying that the algorithm can be scaled up and allow several MTFSs to be calibrated simultaneously. The calibration curve (Fig. 3c) exhibits nonlinearity which is expected for the thermal type of flow sensors – the higher the flow rate, the lower the sensor sensitivity. To ensure the MTFS accuracy is in line with the reference thermal FS's, the MTFS working range was set to 2-30 µL/min. According to experimental measurements, the proposed MTFS allows to measure fluid flow in the range of 2-30 µL/min with a relative error of less than 5% of the measured value (MV) compared to the reference Coriolis sensor. The absolute error is less than 1 µL/min at the maximum flow rate of 30 µL/min and less than 0.4 µl/min at a near-zero flow rate (Fig. 3d and 3e).

To define the response time of the MTFS to the rapid flow alterations, a piecewise meander pressure function was set at the channel inlet. During the experiment, the time intervals each sensor needed to indicate the flow rate increase from 0 to 90% of the full range ($T_1$) and the flow rate decrease from 100 to 10% of the full range ($T_2$), were measured (Fig. 3f). Compared to the reference thermal FS, the response times $T_1$ and $T_2$ of the proposed MTFS were 9% and 12% shorter, respectively. Moreover, the reference thermal FS tended to overestimate the actual value just after a sharp flow rate increase. The reference Coriolis FS had shorter $T_1$ and $T_2$ than thermal flow sensors, but it demonstrated pronounced graph peaks overshooting by more than 20% after flow the rate up or down steps. This shortcoming should be minded in applications where rapid alteration of flow rate is needed. In contrast to both reference FSs, the proposed MTFS demonstrated outstanding measurement stability and fast response without the described shortcoming. The proposed MTFS was extensively tested and demonstrated characteristics in line with commercial devices' (Table S2).

To summarize, the developed MTFS feature high measurement accuracy, performance stability and low response time for rapidly changing flows in low range. The developed device is one of the basic elements for a complex LOC and is used for flow control on a microfluidic chip. An unprecedented opportunity of the MTFS integration into a microfluidic chip makes the sensor essential for highly accurate dosing, sample dilution and flow control on a multichannel LOC.

*December 21, 2022*

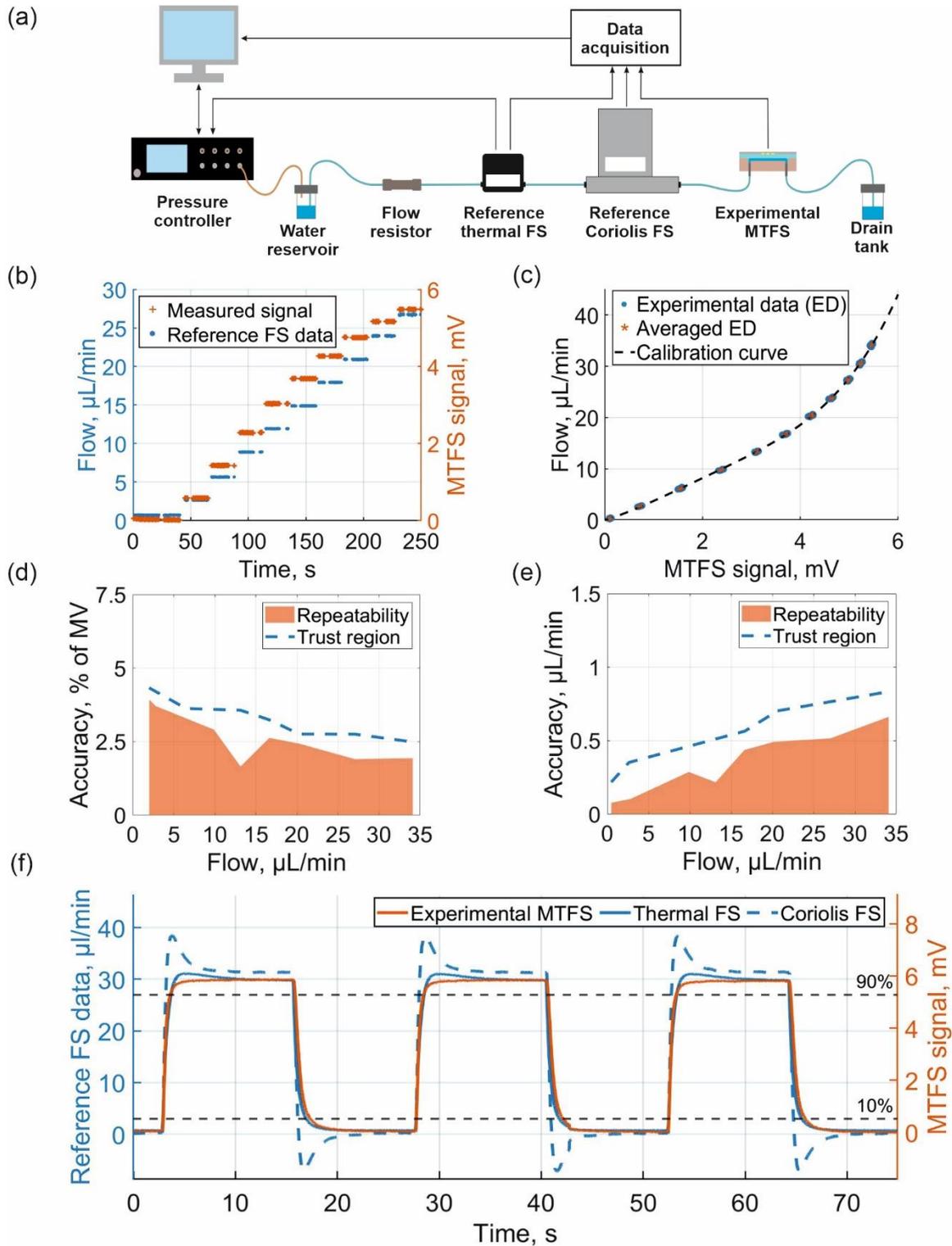

Figure 3 – (a) experimental setup scheme; (b) filtered data from the reference sensor and the MTFS; (c) The MTFS calibration curve; (d) absolute and (e) relative accuracy of the MTFS; (f) MTFS and two reference sensors responses on stepwise flow rate change.

*December 21, 2022*

## 4. Conclusion

Multivariate design optimization, wafer-scale fabrication and automated calibration of a chip-integrated thermal flow sensor for advanced silicon-glass microfluidic devices were presented. The simulation resulted in strict rules for sensing elements' topology and channel geometry. The technology involved both microelectronic (photolithography, electron-beam deposition) and MEMS (DRIE, anodic bonding) processes, and was implemented on 4" wafers. During the tests, the developed sensor demonstrated characteristics in line with or better than those of current commercially available flow sensors.

The MTFS is fully biocompatible and capable of measuring flow rates in the range of 2-30 μL/min with an error of less than 5%. The key feature of the MTFS is its construction. Owing to the membrane-free design and external location of the sensing elements, the MTFS can withstand aggressive chemicals and measure the flow of corrosive reagents during on-chip biological analysis or synthesis. To improve the MTFS characteristics, the switch to a 100 μm thick glass wafer is definitely necessary.

Further development of on-chip functional elements for chemical sensing and flow actuation will be carried out in the future, to organize a portable standalone microfluidic system. We believe that the developed MTFS will become an important functional unit of microfluidic devices for new-generation analytical systems and will bring such transformative technologies as the human body-on-a-chip towards real-world use.

## Acknowledgements

The research was performed at the BMSTU Nanofabrication Facility (Functional Micro/Nanosystems, FMNS REC, ID 74300).

*December 21, 2022*

The authors acknowledge the whole team of the FMN Laboratory, especially Michail Andronic and Alexander Baburin for fruitful discussions, as well as Olya Efremova for being an example of kindness for all of us.

**Author contributions**

I.A.Rod. and A.V.Z. conceived the device. A.V.Z. and V.V.R. developed the device design. I.A.Rod., I.A.Ryz., A.V.Z. and V.V.R. developed the device fabrication route. T.K., D.A.B., E.S.L., R.Sh.A., A.K.K., E.S.K., V.V.E. and A.V.Z. fabricated and assembled the devices. V.V.R. designed and developed software for device calibration. A.V.Z. and V.V.E. developed electronics for standalone device. V.V.R., A.V.Z., V.V.E., E.S.L. conducted the experiments. I.A.Rod., I.A.Ryz., A.V.Z., V.V.R. analyzed the results. V.V.R. developed the simulation. A.R.A., R.Sh.A., P.G.R. and V.V.R. prepared figures. V.V.R., A.R.A. and A.A.B. wrote draft of the manuscript. All authors read, revised, and approved the final manuscript. I.A.Rod. supervised the project.

**Competing interests**

The authors declare no competing interests.

**References**

[1]  Schulte, T. H., Bardell, R. L., & Weigl, B. H. (2002). Microfluidic technologies in clinical diagnostics. Clinica Chimica Acta, 321(1-2), 1-10.

[2]  Ayuso, J. M., Virumbrales-Muñoz, M., Lang, J. M, Beebe, D. J. (2022). A role for microfluidic systems in precision medicine. Nature Communications, 13, 3086.

[3]  Shin, D. J., Andini, N., Hsieh, K., Yang, S., Wang, T. (2019). Emerging Analytical

Techniques for Rapid Pathogen Identification and Susceptibility Testing. The Annual Review of Analytical Chemistry, 12(1), 41–67.

[4]  Whitesides, G. M. (2006). The origins and the future of microfluidics. Nature, 442(7101), 368-373.

[5]  Roman, G. T., Chen, Y., Viberg, P., Culbertson, A. H., Culbertson, C. T. (2007). Single-cell manipulation and analysis using microfluidic devices. Analytical and Bioanalytical Chemistry, 387(1), 9–12.

[6]  Temiz, Y., Delamarche, E. (2018). Sub-nanoliter, real-time flow monitoring in microfluidic chips using a portable device and smartphone. Scientific Reports, 8(1), 10603.

[7]  Luo, T., Fan, L., Zhu, R., Sun, D. (2019). Microfluidic Single-Cell Manipulation and Analysis: Methods and Applications. Micromachines, 10(2), 104.

[8]  Zhao, C., Ge, Z., Yang, C. (2017). Microfluidic Techniques for Analytes Concentration. Micromachines, 8(1), 28.

[9]  Wang, X., Yi, L., Mukhitov, N., Schrell, A. M., Dhumpa, R., & Roper, M. G. (2015). Microfluidics-to-mass spectrometry: a review of coupling methods and applications. Journal of Chromatography A, 1382, 98-116.

[10] Baburin, Aleksandr S., et al. "Crystalline structure dependence on optical properties of silver thin film over time." 2017 Progress in Electromagnetics Research Symposium-Spring (PIERS), 1497-1502. IEEE, 2017.

[11] Yankovskii, G. M., et al. "Structural and optical properties of single and bilayer silver and gold films." Physics of the Solid State 58.12 (2016): 2503-2510.

[12] Baburin, Alexander S., et al. "Highly directional plasmonic nanolaser based on high-performance noble metal film photonic crystal." Nanophotonics VII Proc. SPIE 10672. 106724D, 2018.

[13] Lake, J. R., Heyde, K. C., Ruder, W. C. (2017). Low-cost feedback-controlled syringe pressure pumps for microfluidics applications. PLOS ONE, 12(4), e0175089.

[14] Li, Z., Mak, S. Y., Sauret, A., Shum, H. C. (2014). Syringe-pump-induced fluctuation in all-aqueous microfluidic system implications for flow rate accuracy. Lab on a Chip, 14(4), 744.

[15] Watson, C., Senyo, S. (2019). All-in-one automated microfluidics control system. HardwareX, 5, e00063.

[16] Lai, X., Yang, M., Wu, H., Li, M. (2022). Modular Microfluidics: Current Status and Future Prospects. Micromachines (Basel), 13(8), 1363.

[17] Kuo, J. T. W., Yu, L., Meng, E. (2012). Micromachined Thermal Flow Sensors—A Review. Micromachines, 3(3), 550–573.

[18] Yu, H., Li, D., Roberts, R. C., Xu, K., & Tien, N. C. (2012). Design, fabrication and testing of a micro-Venturi tube for fluid manipulation in a microfluidic system. Journal of Micromechanics and Microengineering, 22(3), 035010.

[19] Sparreboom, W., Van de Geest, J., Katerberg, M., Postma, F., Haneveld, J., Groenesteijn, J., … & Lötters, J. (2013). Compact mass flow meter based on a micro Coriolis flow sensor. Micromachines, 4(1), 22-33.

[20] Chen, X., Liu, C., Yang, D., Liu, X., Hu, L., & Xie, J. (2019). Highly accurate airflow volumetric flowmeters via pMUTs arrays based on transit time. Journal of

Microelectromechanical Systems, 28(4), 707-716.

[21] Yang, Y., Wang, D., Niu, P., Liu, M., & Zhang, C. (2019). Measurement of vertical gas-liquid two-phase flow by electromagnetic flowmeter and image processing based on the phase-isolation. Experimental Thermal and Fluid Science, 101, 87-100.

[22] Baker, R. C. (1993). Turbine flowmeters: II. Theoretical and experimental published information. Flow measurement and Instrumentation, 4(3), 123-144.

[23] Brody, W. R., & Meindl, J. D. (1974). Theoretical analysis of the CW Doppler ultrasonic flowmeter. IEEE transactions on biomedical engineering, (3), 183-192.

[24] Ryzhkov, V. V., Zverev, A. V., Andronik, M., Echeistov, V. V., Issabayeva, Z. H., Sorokina, O. S., Konstantinova, T., Lotkov, E. S., Ryzhikov, I. A., Rodionov, I. A. (2020). Integrated microfluidic flow sensor for lab-on-chip and point-of-care applications. Biotekhnologiya, 36(4), 112.

[25] Schöler, L., Lange, B., Seibel, K., Schäfer, H., Walder, M., Friedrich, N., ... & Böhm, M. (2005). Monolithically integrated micro flow sensor for lab-on-chip applications. Microelectronic Engineering, 78-79, 164–170.

[26] Doh, I., Sim, D., & Kim, S. S. (2022). Microfluidic Thermal Flowmeters for Drug Injection Monitoring. Sensors, 22(9), 3151.

[27] Xu, W., Lijin, B., Duan, M., Wang, X., Wicaksana, J., Min, A., ... & Lee, Y. K. (2018, January). A wireless dual-mode micro thermal flow sensor system with extended flow range by using CMOS-MEMS process. In *2018 IEEE Micro Electro Mechanical Systems (MEMS)* (pp. 824-827). IEEE.

[28] Hoera, C., Skadell, M. M., Pfeiffer, S. A., Pahl, M., Shu, Z., Beckert, E., & Belder, D. (2016). A chip-integrated highly variable thermal flow rate sensor. Sensors and Actuators B: Chemical, 225, 42–49.

[29] Kim, J., Cho, H., Han, S. I., Han, A., & Han, K. H. (2019). A disposable microfluidic flow sensor with a reusable sensing substrate. Sensors and Actuators B: Chemical, 288, 147-154.

[30] Ashauer, M., Glosch, H., Hedrich, F., Hey, N., Sandmaier, H., & Lang, W. (1999). Thermal flow sensor for liquids and gases based on combinations of two principles. Sensors and Actuators A: Physical, 73(1-2), 7–13.

[31] Elwenspoek, M. (1999, October). Thermal flow micro sensors. In CAS'99 Proceedings. 1999 International Semiconductor Conference (Cat. No. 99TH8389) (Vol. 2, pp. 423-435). IEEE.

[32] Xia, Y., & Whitesides, G. M. (1998). Soft lithography. Angewandte Chemie International Edition, 37(5), 550-575.